\documentstyle[mnextra]{mn} 
\input{epsf}
\nobrackets     

\def\gsim{\mathrel{\raise0.35ex\hbox{$\scriptstyle >$}\kern-0.6em
\lower0.40ex\hbox{{$\scriptstyle \sim$}}}}
\def\lsim{\mathrel{\raise0.35ex\hbox{$\scriptstyle <$}\kern-0.6em
\lower0.40ex\hbox{{$\scriptstyle \sim$}}}}
\def\gs{\mathrel{\raise0.35ex\hbox{$\scriptstyle >$}\kern-0.6em
\lower0.40ex\hbox{{$\scriptstyle \sim$}}}}
\def\ls{\mathrel{\raise0.35ex\hbox{$\scriptstyle <$}\kern-0.6em
\lower0.40ex\hbox{{$\scriptstyle \sim$}}}}

\def\kms{\,\hbox{km}\,\hbox{s}^{-1}}
\def\Msol{\mathrel{\rm M_{\odot}}}
\def\Wm2{\,\hbox{W}\,\hbox{m}^{-2}}

\begin{document}

\title[Optical/Near-IR Integral Field Spectroscopy of N2-850.4]{Optical
  and Near-Infrared Integral Field Spectroscopy of the SCUBA Galaxy
  N2-850.4}

\author[Swinbank et al.]{
\parbox[h]{\textwidth}{
A.\,M.\ Swinbank,$^{1\, *}$
Ian Smail,$^1$
R.\,G.\ Bower,$^1$
C.\,Borys,$^2$
S.\,C.\ Chapman,$^2$
A.\,W.\ Blain,$^2$
R.\,J.\ Ivison,$^{3,4}$
S.\, Ramsay Howat,$^3$
W.\,C. Keel$^5$
\& A.\,J. Bunker\,$^6$}
\vspace*{6pt} \\
$^1$Institute for Computational Cosmology, Department of
Physics, University of Durham, South Road, Durham DH1 3LE, UK \\
$^2$Astronomy Department, California Institute of
Technology, 105-24, Pasadena, CA 91125, USA \\
$^3$Astronomy Technology Centre, Royal Observatory,
Blackford Hill, Edinburgh, EH19 3HJ, UK \\
$^4$Institute for Astronomy, University of Edinburgh,
Edinburgh,EH19 3HJ, UK \\
$^5$Department of Physics and Astronomy, University of Alabama,
Tuscaloosa, AL 35487, USA \\
$^6$Department of Physcis, University of Exeter, Stocker Road, Exeter, EX4
4QL, UK \\
$^*$Email: a.m.swinbank@durham.ac.uk \\
}

\maketitle

\begin{abstract}
  We present optical and near-infrared integral field spectroscopy of
  the SCUBA galaxy SMM\,J163650.43+405734.5 (ELAIS N2 850.4) at
  z=2.385.  We combine Ly$\alpha$ and H$\alpha$ emission line maps and
  velocity structure with high resolution {\it HST} ACS and NICMOS
  imaging to probe the complex dynamics of this vigorous star-burst
  galaxy.  The imaging data shows a complex morphology, consisting of
  at least three components separated by $\sim$1$''$ (8\,kpc) in
  projection.  When combined with the H$\alpha$ velocity field from
  UKIRT UIST IFU observations we identify two components whose
  redshifts are coincident with the systemic redshift, measured from
  previous CO observations, one of which shows signs of AGN activity.
  A third component is offset by $220\pm50$$\kms$ from the systemic
  velocity.  The total star formation rate of the whole system
  (estimated from the narrow-line H$\alpha$ and uncorrected for
  reddening) is 340$\pm$50$\Msol$${\rm yr^{-1}}$.  The Ly$\alpha$
  emission mapped by the GMOS IFU covers the complete galaxy and is
  offset by $+270\pm40\kms$ from the systemic velocity.  This velocity
  offset is comparable to that seen in rest-frame UV-selected galaxies
  at similar redshifts and usually interpreted as a star-burst driven
  wind.  The extended structure of the Ly$\alpha$ emission suggests
  that this wind is not a nuclear phenomenon, but is instead a galactic
  scale outflow.  Our observations suggest that the vigorous activity
  in N2 850.4 is arising as a result of an interaction between at least
  two dynamically-distinct components, resulting in a strong starburst,
  a starburst-driven wind and actively-fuelled AGN activity.  Whilst
  these observations are based on a single object, our results clearly
  show the power of combining optical and near-infrared integral field
  spectroscopy to probe the power sources, masses and metallicities of
  far-infrared luminous galaxies, as well as understanding the role of
  AGN- and star-burst driven feedback processes in these high redshift
  systems.
\end{abstract}


\begin{figure*}
\epsfxsize=7.0 truein \epsfbox{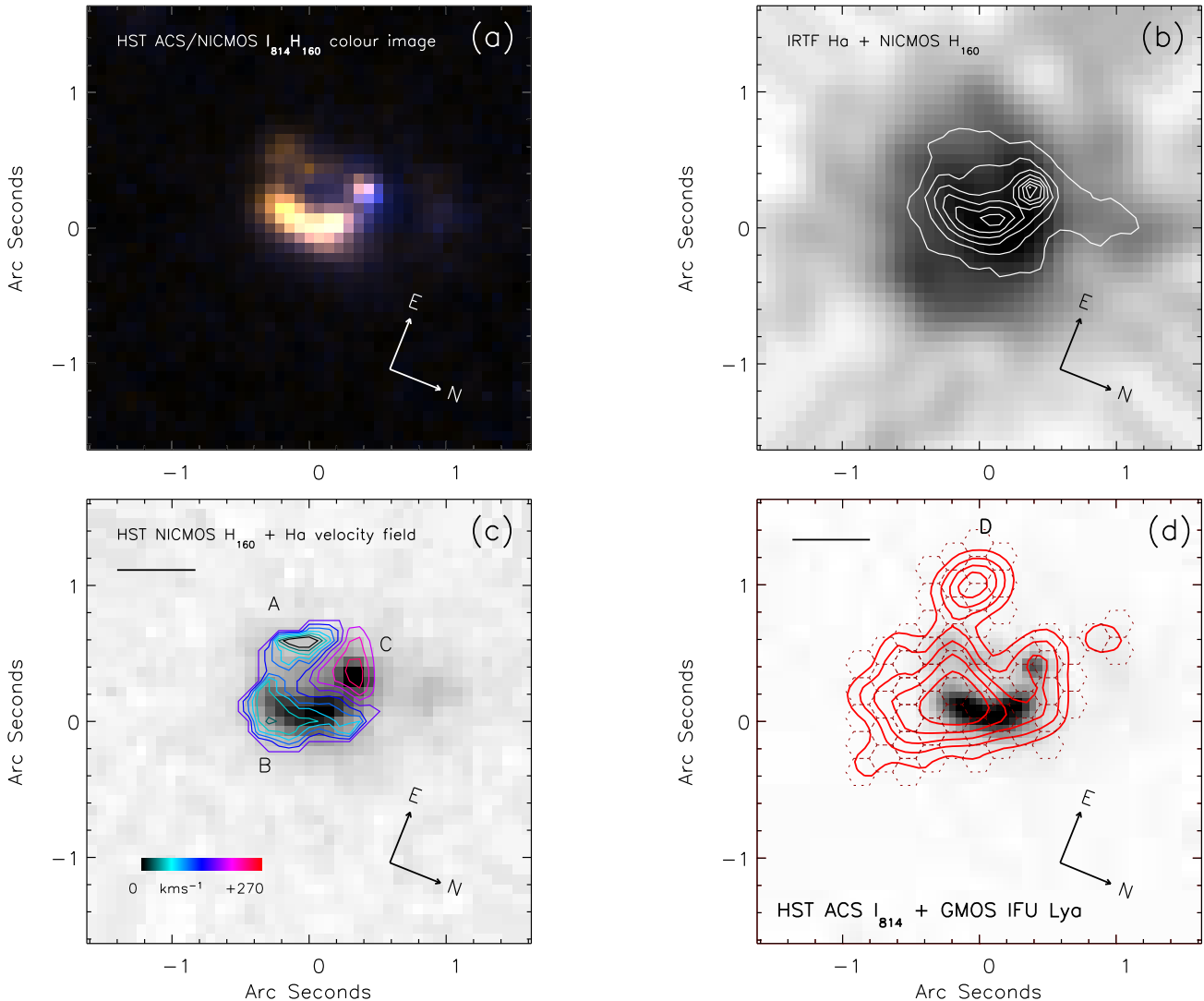}
\caption{\small (a) True colour $I_{814}H_{160}$ image of 
  N2 850.4 from the {\it HST} ACS and NICMOS imaging.  The image shows
  a complex morphology, with at least three distinct components
  separated by $\sim1''$ ($\sim$8\,kpc) in projection.  (b) IRTF
  H$\alpha$ narrow-band image of N2 850.4 with the contours from the
  NICMOS $H_{160}$-band image overlaid.  This H$\alpha$ narrow-band
  image shows a diffuse halo of material distributed asymmetrically
  around the galaxy [seeing $\sim$0.7$''$].  (c) The velocity field of
  N2 850.4 derived from UIST IFU observations of the H$\alpha$ emission
  line overlaid on the NICMOS $H_{160}$-band image.  The redshift of
  component {\it A} is in excellent agreement with previous CO
  observations (Neri et al.\ 2003).  Components {\it A} and {\it B} are
  separated by $50\pm50\kms$ whilst there is a velocity difference of
  $+270\pm50\kms$ between components $A$ and $C$ [the 0.6$''$ seeing is
  marked by the solid bar in the top left hand corner].  (d) {\it HST}
  ACS $I_{814}$-band (F814) image of N2 850.4 with the Ly$\alpha$
  intensity from the GMOS IFU overlaid as contours (the contours mark
  3,4,5,6 and 7$\sigma$).  We also overlay the footprint of the GMOS
  IFU fibers which have $>3\sigma$ emission line detections.  The solid
  bar in the top left hand corner of this panel represent 0.6$''$
  seeing.  The Ly$\alpha$ contours match the high surface brightness
  emission traced by the $I_{814}$-band imaging data, although there is
  Ly$\alpha$ emission to the East (labelled {\it D}) which is not seen
  in the $I_{814}-$band morphology.  The Ly$\alpha$ is redshifted from
  the systemic by $+270\pm40\kms$ which may be indicative of a
  galactic-scale outflow.}
\end{figure*}

\section{Introduction}
Recent surveys have concluded that a substantial fraction of the high
redshift submillimetre (sub-mm) selected galaxy population comprises
morphologically complex systems with high instantaneous star formation
rates and actively fuelled AGN (Smail, Ivison \& Blain 1997; Barger et
al.\ 1999; Scott et al.\ 2002; Smail et al.\ 2002; Ivison et al.\ 2002;
Alexander et al.\ 2003, 2005; Chapman et al.\ 2003; Webb et al.\ 2003;
Dannebauer et al.\ 2004; Knudsen 2004; Swinbank et al.\ 2004; Chapman
et al.\ 2005a; Pope et al.\ 2005).  Understanding the importance of
this population requires identifying their power source (e.g.\, to
determine whether star formation- or AGN- activity dominate the
luminosity output) and, perhaps more importantly, masses for these
galaxies.  Although they have only moderate space densities, their
apparantly high star formation rates mean their contribution to the
cosmic star formation rate could be significant (e.g.\, Chapman et al.\ 
2005b).  Moreover, the star formation activity suggests {\it a priori}
that these galaxies should house starburst-driven superwinds --
outflows which expel gas from the galaxy potential (e.g.\, Pettini et
al.\ 2001; Shapley et al.\ 2003) and which are believed to play an
important role in regulating galaxy formation, preventing the bulk of
baryons cooling into stars (the ``cosmic cooling crisis'', White \&
Rees 1978; Cen \& Ostriker 1999; Balogh et al.\ 2001).  However, to
study the energetics and dynamics of these frequently complex systems
(e.g.\, Smail et al.\ 2004) we must trace the distribution of the
velocity and intensity of emission lines on sub-arcsecond scales.
Ideally, this should be achieved in 2-D to untangle the complex
morphologies of these systems and, in addition to search for signatures
of lensing, which might provide a more mundane explanation of the
apparently intense luminosities of these galaxies (Tecza et al.\ 2004).

In this paper we demonstrate the power of combining optical and
near-infrared integral field spectroscopy with high resolution {\it
  Hubble Space Telescope (HST)} imaging to study the dynamics,
morphologies, masses and outflows of SCUBA galaxies.  Using the
Integral Field Units (IFUs) on GMOS (optical) and UIST (near-infrared)
we have studied the SCUBA galaxy SMM\,J163650.43+405734.5 (ELAIS N2
850.4; Scott et al.\ 2002; Ivison et al.\ 2002; Smail et al.\ 2003).
In \S2 we present the data reduction and results from the spectroscopic
and imaging data. In \S3 and \S4 we present our analysis and
conclusions respectively.  We use a cosmology with $H_{0}=70\kms$,
$\Omega_{M}=0.3$ and $\Omega_{\Lambda}=0.7$ in which 1$''$ corresponds
to 8.2\,kpc at $z=2.4$.

\section{Observations and Analysis}
N2 850.4 was first catalogued as a bright ($8.2\pm1.7$mJy) sub-mm
source by Scott et al.\ (2002) and identified through its radio
counterpart by Ivison et al.\ (2002).  A spectroscopic redshift of
$z=2.38$ for the radio counterpart was measured by Chapman et al.\ 
(2003, 2005a).  N2 850.4 has a far-infrared bolometric luminosity of
L$_{\rm FIR}$=3.1$\times$10$^{13}$L$_{\odot}$ (Chapman et al.\ 2005a)
which corresponds to a star formation rate of
$\sim$5400$\Msol$yr$^{-1}$, (Kennicutt 1998); (although the
far-infrared luminosity may have a contribution from a non-thermal
(AGN) component).  Interferometric observations of the molecular CO
emission in this system by Neri et al.\ (2003) have tied down the
systemic redshift as $z=2.384\pm0.001$ and indicate a gas mass of
$5.5\times10^{10}\Msol$.  The system was studied in detail by Smail et
al.\ (2003) whose multi-wavelength longslit observations suggest that
this system comprises at least two components, one of which has a
Seyfert-like AGN and the other maybe a UV-bright starburst with an
outflow.  These observations show extended [O{\sc iii}]$\lambda$5007
emission as well as strong UV stellar absorption features.

However, due to the multi-component nature of this system and the way
in which longslit observations mix spatial and spectral resolution, the
observations of this galaxy have been difficult to interpret (Smail et
al.\ 2003).  By targeting N2 850.4 with an IFU we are able to decouple
the spatial and spectral resolution and cleanly probe the dynamics and
power sources of this hyper-luminous SCUBA galaxy.

\subsection {HST Optical and Near-Infrared Imaging}

{\it HST} Advanced Camera for Surveys (ACS) observations were obtained
from the {\it HST} public archive\footnotemark (Program ID \#9761).
The data consist of dithered exposures with the F814W filter, taken in
{\sc lowsky} conditions using the default four-point {\sc
  acs-wfc-dither-box} configuration. This pattern ensures optimal
half-pixel sampling along both coordinates. The total integration time
was 4.8\,ks.  We reduced the data using the latest version of the {\sc
  multidrizzle} software (Koekemoer et al. 2002) using the default
parameters with {\sc pixfrac=1} and {\sc scale=1}.  The resulting image
has 0.05$''$ pixels and is free from artifacts (Fig.~1).

\footnotetext{Obtained from the Multimission Archive at the Space
  Telescope Science Institute (MAST).  STScI is operated by the
  Association of Universities for Research in Astronomy, Inc., under
  NASA contract NAS5-26555. Support for MAST for non-{\it HST} data is
  provided by the NASA Office of Space Science via grant NAG5-7584 and
  by other grants and contracts.}
  
The NICMOS data were obtained in Cycle 12, and the target was observed
using the NIC2 camera in the F160W filter for a total of 2.3\,ks
(Program ID \#9856).  We employed the standard four point spiral dither
pattern, {\sc lowsky} conditions and used the {\sc multiaccum}
readmode.  Each exposure was corrected for a pedastal offset, and then
mosaiced using the {\sc calnicb} task in {\sc iraf}.  Unfortunately the
observation was effected by the South Atlantic Anomaly (SAA), and extra
processing steps were required\footnote{For a full description, see \\
  http://www.stsci.edu/hst/nicmos/tools/post\_SAA\_tools.html}.  The
final images appear very flat and have very low cosmic ray
contamination.  Absolute astrometry of the NICMOS images is accurate to
only $\leq$2$''$, so we cross-correlated the full image against the
high resolution ACS data to align the near-infrared image with the
optical image.  Both are aligned to the FK5 coordinate system of our
deep radio map of this field (Ivison et al.\ 2002) which has an
absolute astrometry precision of 0.3$''$.  A complete discussion of the
optical and near-infrared observations and data-reduction is given in
Borys et al.\ (2005).

We degrade the {\it HST} ACS data to the same resolution as the NICMOS
observations and make a true colour ($I_{814}H_{160}$) image of N2
850.4.  An inspection of this imaging data (Fig.~1) reveals a complex
system made up of several components.  In particular, the brightest
features have similar colours and an apparent geometry which is
reminiscent of a triply-imaged, strongly lensed system.  Could the
immense luminosity of N2 850.4 be due to strong lensing (e.g.\ Chapman
et al.\ 2002)?.  Our IFU observations of this system provide a powerful
tool for testing this suggestion, since the redshifts and spectral
features should be the same for all three components if they are all
images of a single background galaxy.

\subsection{IRTF Narrow-band Imaging}

Narrow-band imaging of N2 850.4 was carried out using the 3-m NASA
Infra-Red Telescope Facility\footnotemark (IRTF) Telescope between 2003
April 28 and May 02.  The observations were made in generally
photometric conditions and $\sim0.7''$ seeing.  We used the NSFCAM
camera (Shure et al.\ 1993) which employs a $256\times256$ InSb
detector at 0.15$''$\,pixel$^{-1}$ to give a 38$''$ field of view
(which probes roughly 300\,kpc at $z\sim 2.4$).  The continuously
variable tunable narrow-band filter (CVF) in NSFCAM provides an $R=90$
passband which was tuned to target the H$\alpha$ emission at the
systemic galaxy redshift measured ($z=2.384$) from CO and UV spectrum
of Neri et al.\ (2003) and Smail et al.\ (2003) respectively.  Shorter,
matched broad-band imaging were interspersed between the narrow-band
exposures to provide continuum subtraction.  The total narrow-band
integration time was 19.8\,ks and the total broad band integration time
was 2.2\,ks. These observations, their reduction and analysis are
discussed in detail in Swinbank et al.\ (2004).

\footnotetext{The Infrared Telescope Facility is operated by the
  University of Hawaii under Cooperative Agreement no.\ NCC 5-538 with
  the National Aeronautics and Space Administration, Office of Space
  Science, Planetary Astronomy Program.}

\subsection{Spectroscopic Imaging}

\subsubsection{UIST Near-Infrared Integral Field Spectroscopy}

Observations of N2 850.4 were made in queue mode with the UKIRT
Image-Spectrometer (UIST) IFU between 2003 March 27 and April 04 in
$<$0.6$''$ seeing and photometric conditions\footnotemark.  The UIST
IFU uses an image slicer and re-imaging mirrors to reformat a square
array of 14-slices of the sky, (each 0.24$''$$\times$0.12$''$) into a
pseudo-longslit.  The resulting field of view is 3.4$''$$\times$6.0$''$
(Ramsay Howat et al.\ 2004).  We used the $HK$ grism which has a
spectral resolution of $\lambda/\Delta\lambda=1000$ and covers a
wavelength range of 1.4--2.4$\mu$m.  Observations were carried out in
the standard ABBA configuration in which we chopped away to sky by
12$''$ to achieve good sky subtraction.  Individual exposures were 240s
seconds and each observing block was 7.2\,ks which was repeated four
times, thus the total integration time was 28.8\,ks.

\footnotetext{The United Kingdom Infrared Telescope is operated by the
  Joint Astronomy Center on behalf on the UK Particle Physics and
  Astronomy Research Council.}

To reduce the data we used the relevant {\sc orac-dr} pipeline
(Cavanagh et al.\ 2003) which sky-subtracts, extracts, wavelength
calibrates, flatfields, and forms the datacube.  To accurately align
and mosaic the four datacubes we created white light (wavelength
collapsed) images around the redshifted H$\alpha$ emission line from
each observing block and used the peak intensity to centroid the
object in the IFU datacube.  We then spatially aligned and co-added
the four individual data-cubes (weighted by H$\alpha$ signal-to-noise)
using {\sc mosaic} in {\sc kappa}.

To search for velocity structure we attempt to identify H$\alpha$
emission on a pixel-by-pixel basis by averaging over a
0.48$''$$\times$0.48$''$ region (4$\times$2 pixels), increasing to
0.6$''$$\times$0.72$''$ (5$\times$3 pixels) if no emission line could
initially be identified.  At $z=2.385$, H$\alpha$$\lambda6562.8$
emission falls at 2.221$\mu$m, which is away from any strong OH
emission or absorption.  We attempt to fit a single Gaussian to the
H$\alpha$ emission line, but also attempt to identify an [N{\sc
  ii}]$\lambda6583$ emission line, only accepting the fit if the
$\chi^{2}$ is significantly better than without the [N{\sc ii}] line.
We checked the wavelength calibration of each IFU pixel by fitting a
nearby sky line with a Gaussian profile.  The errors in the velocity
field are calculated by building two independent data-cubes, each of
14.4\,ks and recomputing the velocity field in an identical manner to
that described above.  Using the same fitting techniques as above we
estimate that the average velocity error to be $\simeq50\kms$.  Spectra
from the three components identified in the UIST IFU observations are
shown in Fig.~2.

To confirm the velocity gradients seen in the UIST IFU data (Fig.~1) we
also obtained a 4.8\,ks exposure around the H$\alpha$ emission lines
with the Keck near-infrared longslit spectrograph.  We aligned the slit
along components {\it A} and {\it B} in Fig.~1 and derive the same
velocity offsets between components as in our the IFU data (the spectra
and line fluxes are shown and discussed in Swinbank et al.\ 2004).

\begin{figure}
  \epsfxsize=3.5 truein \epsfbox{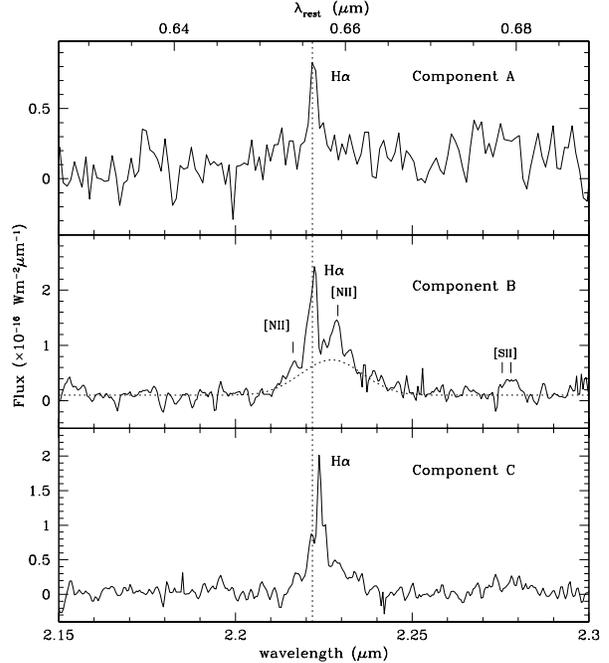}
\caption{\small {\it Top:} Near-infrared spectrum of component 
  {\it A} around the H$\alpha$ emission line.  This component is at the
  same redshift as the systemic redshift ($z=2.384$) as measured from
  the molecular CO emission by Neri et al.\ (2003).  {\it Middle:}
  Near-infrared spectrum of component {\it B} from which shows a
  $+50\pm50\kms$ velocity shift from the systemic.  This component also
  has a broad H$\alpha$ regions offset by $+800\pm150\kms$ which
  indicates AGN activity.  The FWHM of the broad line is
  $2300\pm250\kms$.  {\it Lower:} Near-infrared spectrum of component
  {\it C} which shows a $220\pm50\kms$ velocity offset from the
  systemic.  Component {\it C} also displays a broad line component
  ($FWHM=1800\pm300\kms$) which is at a similar redshift to the broad
  line seen in component {\it B} and may arise due to the same
  scattering seen in [O{\sc iii}]$\lambda$5007 seen by Smail et al.\ 
  (2003).  The top panel has been binned by a factor of two in the
  spectral direction to enhance the contrast of the H$\alpha$ emission.
  The dotted line shows the wavelength for H$\alpha$ expected at the
  systemic redshift of 2.384 (Neri et al.\ 2003; Smail et al.\ 2003).}
\end{figure}

\subsubsection{GMOS Optical Integral Field Spectroscopy}

N2 850.4 was observed with the GMOS-IFU on Gemini North on 2002 June 12
during Science Demonstration time for a total of 7.2\,ks in $0.6''$
seeing and photometric conditions\footnotemark.  The GMOS IFU uses a
lensed fibre system to reformat a $7'' \times 5''$ field into two long
slits (Allington-Smith et al.\ 2002).  Using a $B$-band filter in
conjunction with the B600 grating results in two tiers of spectra
recording the maximum field of view.  The spectral resolution of this
configuration is $\lambda/\Delta\lambda_{\rm FWHM}=2000$.  For the
galaxy at $z=2.385$, the Ly$\alpha$ emission falls at 4112\AA\ which is
in a region of low sky emission, but also low throughput.

\footnotetext{Programme ID: GN-2002A-DD-4.  Obtained at the Gemini
  Observatory, which is operated by the Association of Universities for
  Research in Astronomy, Inc., under a cooperative agreement with the
  NSF on behalf of the Gemini partnership: the National Science
  Foundation (United States), the Particle Physics and Astronomy
  Research Council (United Kingdom), the National Research Council
  (Canada), CONICYT (Chile), the Australian Research Council
  (Australia), CNPq (Brazil) and CONICET (Argentina)}

The GMOS data reduction pipeline was modified such that the extracted
spectra included the blue edge of the CCD (where the Ly$\alpha$
emission falls) and then used to extract and wavelength calibrate the
spectra of each IFU element.  The variations in fibre-to-fibre response
were removed using the continuum around the expected range of
Ly$\alpha$ emission.  To check the wavelength calibration around
4100\AA\, we wavelength calibrated the CuAr arc observations and fit
the arc lines between 4000 and 4200\AA\ with Gaussian profiles.  We
measure the rms offset between the observed arc line centroids and the
arc line list to be less than 0.02\AA\ (which corresponds to less than
8$\kms$ in the rest frame of the galaxy).  This gives us confidence
that any velocity structures or offsets in the GMOS IFU data are real
and not simply an artifact of the observations.  To search for velocity
structure the spectra were averaged over a $3 \times 3$ pixel
(0.6$''$$\times$0.6$''$) spatial region, except where the signal was
too low to give a significant detection of the line, in which case the
smoothing area was increased to $4 \times 4$ pixels.  In regions where
this averaging process still failed to give an adequate $\chi^2$ (i.e.
the inclusion of an emission line component does not improve the fit),
no fit was made.  In order to detect and fit the line we required a
minimum S/N of 3 and checked every fit by eye.  In the inner regions of
the galaxy all of the Gaussian profile fits are accepted, while in the
outer regions we reject fits if the line centroid is greater than
3000$\kms$ away from the systemic, or the best fit Gaussian profile has
a width greater than FWHM$>$3000$\kms$.  Although the Ly$\alpha$
emission is resolved (FWHM$\sim700\kms$), we detect no significant
coherent velocity gradient across the system and place a limit of
$100\kms$ on possible velocity structure (Table~1.).  We construct a
{\it Ly$\alpha$ intensity map} from the emission line and overlay this
on the {\it HST} ACS image in Fig.~1.

\section{Analysis}

To spatially align the imaging and spectroscopy observations we begin
by constructing a H$\alpha$ image from the UIST IFU and align this with
the IRFT H$\alpha$ and continuum images (which are also aligned to the
NICMOS and ACS images using stars in the field of view).  Furthemore,
to tie these to the GMOS data, we construct a Ly$\alpha$ and continuum
image from the GMOS IFU and align these with the galaxy in an observed
$V$-band image from Ivison et al.\ (2002).  This $V$-band image is then
aligned to the near-infrared imaging and results accurate alignment
between the GMOS, UIST, IRTF and HST observations.  We estimate that
the uncertainty in the astrometry between any two frame to be
$\lsim0.2''$.  Having combined the {\it HST } $H_{160}$ and
$I_{814}$-band NICMOS and ACS imaging with the velocity structure of
the H$\alpha$ emission, we find at least three dynamically distinct
components (labelled {\it A}, {\it B}, and {\it C} in Fig.~1) and we
show the near-infrared spectra around the H$\alpha$ emission from these
components in Fig.~2.  The redshift of components {\it A} and {\it B}
are in excellent agreement with previous CO(4--3) observations which
measured the systemic redshift to be $2.384\pm0.001$ (Neri et al.\ 
2003; Greve et al.\ 2005).  The other component, labelled {\it C} is
dynamically distinct from the systemic redshift.  Component {\it B} has
an [N{\sc ii}]/H$\alpha$ emission line ratio of $0.37\pm0.05$, which is
indicative of star-formation, although the presence of an underlying
($2300\pm250\kms$) H$\alpha$ broad line region suggests AGN activity or
scattered light from an AGN.  The velocity offset of the narrow line
H$\alpha$ from the systemic galaxy is $+50\pm50\kms$ and has a width of
$360\pm25\kms$ whilst the broad line H$\alpha$ is redshifted by
$+800\pm150\kms$ (all line widths are deconvolved for the instrumental
resolution).

Turning to the Ly$\alpha$ emission line map from the GMOS IFU
observations (Fig.~1), we find an an extended, diffuse Ly$\alpha$ halo.
Whilst the Ly$\alpha$ seems to roughly follow the $I_{814}$-band
morphology, we also identify Ly$\alpha$ emission lying outside the
optical extent of the galaxy (labelled {\it D} in Fig.~1).  The spatial
extent of the Ly$\alpha$ is $\sim16$\,kpc, however, most interestingly
the velocity of the emission line is placed $+270\pm40\kms$ redward of
the systemic velocity of this system.  We detect no significant
velocity structure in the Ly$\alpha$ emission across the system (see
Table~1).

Using a deep, 1.44$''$ resolution 1.4-GHz map of the field from the
VLA, Ivison et al.\ (2002) identified a compact radio source with a
centroid which corresponds exactly to the location of component {\it B}
in Fig.~1.  This gives us confidence that this component is responsible
for the far-infrared activity.  The third component (labelled {\it C})
is offset from {\it A/B} by $+220\pm50\kms$ and has an upper limit of
[N{\sc ii}]/H$\alpha$$\lsim0.05$ and a width of $320\pm30\kms$.  This
indicates star-formation rather than AGN activity, although there is
also evidence for a broad-line H$\alpha$ component which is at the same
redshift as the broad line seen in component {\it B} (Fig.~2).  This
broad line H$\alpha$ may arise as part of the same scattered emission
seen in the [O{\sc iii}]$\lambda$5007 emission in Smail et al.\ (2003).
To attempt to identify which component hosts the AGN activity, we
construct a (wavelength collapsed) white light image from the datacube
between 2.23$\mu$m and 2.25$\mu$m (i.e.\ the broad-line H$\alpha$
emission) and compare this with the white light image generated by
collapsing the datacube between 2.215$\mu$m and 2.225$\mu$m (which
includes the narrow-line H$\alpha$ emission).  Unfortunately these two
images look very similar and it is not possible to identify which
component hosts the AGN.

The velocity offsets and spectral differences seen among the various
morphological components immediately rules out the possibility that all
are gravitationall lensed images of a single background source.
Instead, it appears that N2-850.4 is a multi-component and complex
merger.  Assuming the velocity offsets arises due to merging components
in the potential well, we estimate a dynamical mass of
$\gsim2\times10^{11}\Msol$.

The velocity offsets from the H$\alpha$ emission can be compared
directly to the dynamics from the CO(4--3) and CO(7--6) observations
from Neri et al. (2003) (see also Greve et al.\ 2005).  It appears that
the broad CO(4--3) emission arises from two components (a bright
component at z=2.383, and a fainter component at slightly higher
redshift (at z$\sim$2.388) which is located $\gsim0.3''$ to the NE of
the first component).  Narrow CO(7--6) emission is also detected at the
same position and redshift as the lower redshift CO(4--3) emission.
Assuming the CO(7--6) and the lower redshift CO(4--3) emission arise
from warm, dense gas associated with {\it A} and/or {\it B} and the
higher redshift CO(4--3) emission arises from component {\it C}, the
velocity and spatial offsets are in excellent agreement with our IFU
observations.  Furthermore, this suggests that {\it A/B} and {\it C}
all host gas reservoirs, with {\it B} being the most massive.  The
implied molecular gas mass from the CO observations is
$\sim5.5\times10^{10}\Msol$, thus the dynamical mass
($\sim2\times10^{10}\Msol$) for N2 850.4 is approximately four times
greater than gas mass estimate suggesting that this system has a high
baryonic fraction in the central regions.

Using the H$\alpha$ emission line as a star-formation rate indicator we
can calculate the star formation rate in each of the three components.
For solar abundances and adopting a Salpeter IMF, the conversion
between H$\alpha$ flux and star formation rate is SFR($\Msol$${\rm
  yr^{-1}}$)=$7.9\times10^{-35}$L(H$\alpha$)$\Wm2$ (Kennicutt 1998).
This calibration assumes that all of the ionising photons are
reprocessed into nebular lines (i.e. they are neither absorbed by dust
before they can ionise the gas, nor do they escape the galaxy).  Using
the narrow-line H$\alpha$ emission line fluxes with this calibration we
find that the star formation rates of components {\it A}, {\it B} and
{\it C} (uncorrected for reddening) are $\lsim30$, 150$\pm$30 and
140$\pm$30$\Msol$${\rm yr^{-1}}$respectively.  The total star formation
rate is a factor of $\gsim10$ less than the star formation rate implied
from the far-infrared luminosity, and implies aproximately three
magnitudes of dust extinction (e.g.\ Smail et al.\ 2004).  \medskip

\begin{table*}
\begin{center}
{\footnotesize
{\centerline{\sc Table 1.}}
{\centerline{\sc Velocities of rest frame UV and optical emission}}
\begin{tabular}{lcccccccc}
\hline
\noalign{\smallskip}
Component   & z$_{H\alpha}$ & FWHM$_{H\alpha}$ & H$\alpha$ Flux & z$_{Ly\alpha}$ & v$_{H\alpha}$ & v$_{Ly\alpha}$ \\
            &               &   ($\kms$)       & ($10^{-19}\Wm2$) &                &    ($\kms$)   &   ($\kms$)     \\
\hline
\hline
A           & 2.3841[4]     &   330$\pm$40     & $0.9\pm0.2$    &    ...         &    0$\pm$50   & ...            \\
B           & 2.3847[3]     &   360$\pm$25     & $4.4\pm0.9$    & 2.3870[4]      &  +50$\pm$50   & +260$\pm$40    \\
B$_{broad}$ & 2.3930[10]    &  2300$\pm$250    & $16.7\pm2.0$   & ...            & +800$\pm$150  & ...            \\
C           & 2.3866[3]     &   320$\pm$30     & $4.0\pm1.0$    & 2.3872[4]      & +220$\pm$50   & +275$\pm$40    \\
C$_{broad}$ & 2.3900[12]    &  1800$\pm$300    & $8.0\pm1.5$    & ...            & +550$\pm$200  & ...            \\
D           &    ...        &   ...            &    ...         & 2.3860[6]      & ...           & +170$\pm$50    \\
\hline
\end{tabular}
\begin{minipage}{4.5in}
  Notes: The value given in the [] $z$ column is the error in the
  last decimal place.  The quoted velocities are with respect to
  the systemic redshift from Neri et al.\ 2003.
\end{minipage}
}
\end{center}
\medskip
\end{table*}

\section{Discussion \& Conclusions}

The colours and morphology of N2 850.4 from our {\it HST} ACS and
NICMOS imaging resemble that of a strongly lensed galaxy, however the
lens interpretation is quickly ruled out from the H$\alpha$ emission
maps which show that this system is made up of at least three
dynamically distinct components separated by $\sim1''$ (8\,kpc) in
projection and up to $220\pm50\kms$ in velocity.  The ground- and
space-based imaging data also shows a diffuse and asymmetric halo of
material surrounding the galaxy.  The H$\alpha$ redshift of components
{\it A/B} in Fig.~1 are in excellent agreement with previous CO and
rest-frame UV longslit observations which have measured the systemic
redshift to be 2.384 (Neri et al.\ 2003; Smail et al.\ 2003; Greve et
al.\ 2005).  The presence of an underlying broad ($\sim2000\kms$)
emission line (offset by $\sim+800\kms$) in components {\it B} and {\it
  C} suggests AGN activity.  Comparable narrow-line to broad-line
velocity offsets are frequently seen in local Seyfert nuclei (e.g.\ 
Ostenbrook \& Shuder 1987; Corbin et al.\ 1996; Storchi-Bergmann et
al.\ 2003) as well as high-redshift radio galaxies (e.g.\ Simpson et
al.\ 1999).  The third component detected in narrow-line H$\alpha$
emission {\it C} is redshifted from the systemic by $220\pm50\kms$.
Combined with the high resolution imaging, the complex morphology and
dynamics of this system suggests a massive merger event which has
presumably triggered a strong, obscured star-burst and AGN activity.

The GMOS IFU observations show that N2 850.4 has an extended halo of
Ly$\alpha$ emission. The Ly$\alpha$ halo has a spatial extent of
$\sim16$\,kpc and is redshifted relative to the systemic velocity by
$+270\pm40\kms$.  It is interesting to compare the Ly$\alpha$ emission
from N2 850.4 with the giant sub-mm detected Ly$\alpha$ haloes LAB1 and
LAB2 in the SSA22 field (Steidel et al.\ 2000; Chapman et al.\ 2001,
2004; Bower et al.\ 2004; Wilman et al.\ 2005). From our observations,
N2 850.4 has a slightly lower integrated Ly$\alpha$ luminosity
($\gsim$3$\times$10$^{43}$erg\,s$^{-1}$) compared to LAB1 and LAB2
(1$\times$10$^{44}$erg\,s$^{-1}$ and 9$\times$10$^{43}$erg\,s$^{-1}$
respectively), but is much more compact (LAB1 and LAB2 have areas of
over 100arcsec$^{2}$ (5000\,kpc$^2$)), however the limiting surface
brightness of the GMOS observations are
$\sim$5$\times$10$^{-16}$erg\,s$^{-1}$cm$^{-2}$arcsec$^{-2}$
(significantly less than the surface brightness limit of LAB1 and LAB2
narrow-band imaging observations), and it is therefore possible that N2
850.4 is surrounded by a large-scale diffuse emission halo below the
sensitivity limit of the GMOS IFU observations.  Nevertheless, the
Ly$\alpha$ emission from N2 850.4 is much more peaked than LAB1 and
therefore this halo may represent a different evolutionary phase.
Unfortunately, due to the very different surface brightnesses of the
systems and the large pixel scale of the SAURON IFU (which has 1.0$''$
fibres), it is very difficult to directly compare the dynamics of the
two systems (Bower et al.\ 2004).  In terms of the wider environment,
it is interesting to note that a second sub-mm detected galaxy has
recently been detected in the same structure as N2 850.4 (at a distance
of 2.5\,Mpc; Chapman et al.\ 2005a).

The observed velocity offset between H$\alpha$ and Ly$\alpha$ emission
is comparable to those seen in the spectra of rest-frame UV selected
galaxies at z$\gsim$2 (e.g.\ Teplitz et al.\ 2000; Pettini et al.\ 
2001; Shapley et al.\ 2003) where they have been attributed to galactic
scale outflows produced by the collective effects of heating and
outflows from supernovae.  In this scenario, the Ly$\alpha$ appears
redshifted due to resonant scattering of photons from the inner surface
of a receding shell of material.  Such flows have been termed
``superwinds'' by analogy to the wind seen in the spectra of local
ultra-luminous infrared galaxies and local star-burst galaxies (e.g.\ 
Martin 2005; Keel 2005).  If such winds can escape from the potential
well, they can carry metals to large distances from the galaxy and
deposit large amounts of energy in the intergalactic medium.  These
winds thus have important consequences for the metal enrichment of the
Universe (Aguire et al.\ 2001) and galaxy formation models (Benson et
al.\ 2003).  The key issue, however, is whether the wind material is
localised within individual {H{\sc ii}} regions (in which case, it may
not escape the galaxy potential), or whether the expelled shell already
envelops the complete galaxy.  The latter interpretation is supported
by Adelberger et al.'s (2003) observation of a small scale
anti-correlation between galaxies and Ly$\alpha$ absorption in QSO
spectra (although as those authors quote, the statistical significance
of this result is modest).  If it can be demonstrated that the
superwind shell had already escaped from the galaxy disk, then it may
have enough energy to escape the gravitational potential to distribute
its energy metals widely across the Universe.

To distinguishing between these scenarios, we must examine the spatial
variation of the velocity offset between H$\alpha$ and Ly$\alpha$. Our
data show no correlation between the emission wavelength of Ly$\alpha$
and the velocity variations clearly seen in H$\alpha$. This argues that
the Ly$\alpha$ emission originates outside the individual components.
If we were seeing the inner surface of a shell located well outside the
galaxy we would expect a negligible velocity shear and indeed our
observations place a limit on the shear of $\lsim100\kms$.

We can also investigate how closely the morphology of the Ly$\alpha$
emission traces the star-forming regions of the galaxy.  While the
Ly$\alpha$ intensity map generally traces the $I_{814}$-band
morphology, the diffuse extension labelled {\it D} in Fig.~1, has no
counterpart in the $I_{814}$-band image.  This component may be a dense
knot in an outflowing shell and would be compatible with a model in
which scattered Ly$\alpha$ photons are observed from the outflowing
shell.

The data presesnted here support that idea that we are seeing wind
material that has already escaped from the galaxy.  It is interesting
to compare the velocity offset in Ly$\alpha$ with the escape velocity
of the galaxy.  Using the dynamical mass estimate
($\sim2\times10^{11}\Msol$) enclosed in a radius of $\sim$8\,kpc we
estimate escape velocity to be $\gsim500\kms$ (assuming a central
concentration of $c=7$ for $z=2.4$); (Navarro, Frenk \& White 1997).
Whilst this escape velocity exceeds that of the outflowing material,
the fate of the outflow will depend on its present location (or
equivalently, its initial velocity).  For example, if the outflow
originated in the galaxy with an initial speed of $\sim270\kms$, then
it will surely rain back down on the galaxy.  However, if, as we have
argued, the shell is currently located outside the galaxy (i.e.\ 
$\gsim$10kpc), then the escape velocity will be a factor of $\sim2$
less, in which case the outflow will probably escape the gravitational
potential and distribute the gas much more widely in the environment.
We also note that the UV-bright starburst in N2 850.4 is still
relatively young ($\sim$10Myr; Smail et al.\ 2003) and therefore this
material may still be accelerating into the inter-galactic medium.
Future observations of a larger sample of these galaxies (at various
evolutionary stages) may yield further information about the origin of
the outflows and the size of the regions which they affect around them
(Geach et al.\ 2005).

Our observations suggest that the vigorous activity in N2 850.4 is
arising due to an interaction between at least two distinct components.
One of these contains warm, dense molecular gas and hosts and AGN,
while the second appeard to be less massive, but still contains
substantial amounts of cold gas.  The resulting gravitational tides
resulted in a starburst and (actively fuelled) AGN activity. This
activity has produced a wind that maybe driving enriched gas out into
the inter-galactic medium.  Whilst these observations are based on a
single galaxy, our results clearly show the power of combining optical
and near-infrared observations to probe the power sources, masses and
feedback processes in high redshift, far-infrared luminous selected
galaxies.  The next step is to generate a statistically useful sample
to gauge the prevalence outflows from these massive galaxies which may
explain the processes which shape the galaxy luminosity function and
explain why only 10\% of baryons cool to form stars.

\section*{acknowledgements}   We are very greatful to the referee, Seb
Oliver for his constructive report which significantly improved the
content and layout of this paper.  We would like to thank Brad
Cavanagh, Peter Draper and Stephen Todd for useful discussions and help
regarding the UIST IFU data reduction pipeline and Watson Varicatt and
Sandy Leggett for observing the target with UIST IFU in UKIRT queue
mode in 2003A.  We would also like to thank Matt Mountain and Jean-Rene
Roy for accepting the GMOS IFU programme for Science Demonstration,
Inger J{\o}rgensen and Kathy Roth for vital assistance in observing the
target with GMOS, and Bryan Miller for useful discussion regarding the
GMOS data reduction pipeline.  We acknowledge Roger Davies and Gerry
Gilmore who were the joint PI's of the GMOS-DDT proposal.  We also
acknowledge useful discussions with Alastair Edge, Jim Geach, David
Gilbank, Thomas Greve, Chris Simpson, Martin Ward and Richard Wilman.
AMS acknowledges support from PPARC, IRS acknowledges support from the
Royal Society, RGB acknowledges a PPARC Senior Fellowship and the Euro
3D Research Training Network.  AWB acknowledges support from NSF
AST-0205937 and the Alfred Sloan Foundation.


\begin{thebibliography}{}

\bibitem[]{} Adelberger, K.~L., Steidel, C.~C., Shapley, A.~E.,
  Pettini, M., 2003, ApJ, 584, 45

\bibitem[]{} Aguirre, A., Hernquist, L., Schaye, J., Katz, N.,
  Weinberg, D.~H.; Gardner, J., 2001 ApJ, 561, 521

\bibitem[]{} Alexander, D.~M., Bauer, F.~E., Brandt, W.~N.,
  Hornschemeier, A.~E., Vignali, C., Garmire, G.~P., Schneider, D.~P.,
  Chartas, G., Gallagher, S.~C., 2003, AJ, 125, 383

\bibitem[]{} Alexander, D.~M., Smail, I., Bauer, F.~E, Chapman, S.~C.,
  Blain, A.~W., Brandt, W.~N., Ivison, R.,~J., 2005, Nature, in press.

\bibitem[]{} Allington-Smith, J.~R., Murray, G., Content, R.,
  Dodsworth, G., Miller, B.~W., J{\o}rgensen, I., Hook, I., Davies, R.
  L., et al., 2002, PASP, 114, 79

\bibitem[]{} Balogh, M~L., Pearce, F.~R.; Bower, R.~G.; Kay, S.~T.,
  2001, MNRAS, 326, 1228

\bibitem[]{} Barger, A.~J., Cowie, L.~L., Smail, I., Ivison, R.~J.,
  Blain, A.~W., Kneib, J.-P., 1999, ApJ, 117, 2656

\bibitem[]{} Benson, A.~J., Bower, R.~G., Frenk, C.~S., Lacey, C.~G.,
  Baugh, C.~M., Cole, S., ApJ, 2003, 599, 38

\bibitem[]{} Borys, C., Chapman, S.~C., Blain, A.~W., Smail, I.,
  Ivison, R.~J., 2005 ApJ, in prep.

\bibitem[]{} Bower, R.~G., Morris, S.~L., Bacon, R., Wilman, R.~J.,
  Sullivan, M., Chapman, S., Davies, R.~L., de Zeeuw, P.~T., Emsellem,
  E. 2004, MNRAS, 351, 63

\bibitem[]{} Cavanagh B., Hirst P., Jenness T., Economou F., Currie
  M.~J., Todd S., Ryder S.~D., 2003, in Astronomical Data Analysis
  Software and Systems XII ASP Conference Series, Vol. 295, 2003 H. E.
  Payne, R. I. Jedrzejewski, \& R.  N. Hook, eds., p.237

\bibitem[]{} Cen, R., \& Ostriker, J.~P., 1999, ApJ, 514, 1

\bibitem[]{} Chapman, S.~C., Smail, I., Ivison, R.~J., Blain, 2001,
  ApJ, 548, L17

\bibitem[]{} Chapman, S.~C., Smail, I., Ivison, R.~J., Blain, 2002,
  MNRAS, 335, L17

\bibitem[]{} Chapman, S.~C., Blain, A.~W., Ivison, R.~J., Smail, I.,
  2003, Nature, 422, 695

\bibitem[]{} Chapman, S.~C., Scott, D., Windhorst, R.~A., Frayer, D.,
  Borys, C., Lewis, G.~F., Ivison, R.~J., 2004, ApJ, 606, 85

\bibitem[]{} Chapman, S.~C., Blain, A.~W., Smail, I., Ivison, R.~J.,
  2005a, ApJ, in press.

\bibitem[]{} Chapman, S.~C., Smail, I., Blain, A.~W., Ivison, R.~J.,
  2005b, ApJ, submitted.

\bibitem[]{} Corbin, M.~R., Boroson, T.~A. 1996 ApJ, 107, 69

\bibitem[]{} Condon, J.~J., Anderson, M.~L., Helou, G.  1991, ApJ,
  376, 95

\bibitem[]{} Dannerbauer, H., Lehnert, M.~D., Lutz, D., Tacconi, L.,
Bertoldi, F., Carilli, C., Genzel, R., Menten, K. M., 2004, ApJ, 606
664

\bibitem[]{} Garnett, M.~A. Shields, G.~A., Skillman, E.~D., Sagan,
  S.~P., Dufour, R.~J., 1997, ApJ, 489, 63

\bibitem[]{} Geach, J., et al.\ 2005, MNRAS, in prep.

\bibitem[]{} Genzel, R., Baker, A.~J.; Tacconi, L.~J., Lutz, D., Cox,
  P., Guilloteau, S., \& Omont, A., 2003, ApJ, 584, 633

\bibitem[]{} Greve, T.~R., et al.\, 2005, MNRAS, submitted

\bibitem[]{} Ivison, R.~J., Greve, T.~R., Smail, Ian, Dunlop, J.~S.,
  Roche, N.~D., Scott, S.~E., Page, M.~J., Stevens, J.~A., Almaini, O.,
  Blain, A.~W., Willott, C.~J., Fox, M.~J., Gilbank, D.~G., Serjeant,
  S., Hughes, D. H., 2002, MNRAS, 337, 1

\bibitem[]{} Greve, T.~R., Ivison, R.~J., Bertoldi, F., Stevens, J.~A.,
  Dunlop, J.~S., Lutz, D., Carilli, C.~L., 2004 MNRAS 354 779

\bibitem[]{} Keel, W.~C. 2005, AJ, submitted

\bibitem[]{} Kennicutt, R.~C. 1998, ARAA, 36, 189
  
\bibitem[]{} Koekemoer A.~M., Fruchter A.~S., Hook R.~N., \& Hack W.,
  2002, Proceedings: The 2002 {\it HST} Calibration Workshop,Baltimore,
  MD, 2002.~ ed. S. Arribas, A.  Koekemoer, B Whitmore.

\bibitem[]{} Knudsen, K., 2004 PhD. Thesis, University of Leiden

\bibitem[]{} Neri, R., Genzel, R., Ivison, R.~J., Bertoldi, F., Blain,
  A.~W., Chapman, S.~C., Cox, P., Greve, T.~R., Omont, A., Frayer,
  D.~T. 2003, ApJ, 597, 113

\bibitem[]{} Martin, C., 2005, ApJ, submitted

\bibitem[]{} Neri, R., Genzel, R., Ivison, R.~J., Bertoldi, F., Blain,
  A.~W., Chapman, S.~C., Cox, P., Greve, T.~R., Omont, A., Frayer, D.
  T., 2003, ApJ, 597, 113

\bibitem[]{} Navarro, Julio F., Frenk, C.~S., White, S.~D.~M. 1997,
  ApJ, 490, 493

\bibitem[]{} Ostenbrook, D.~E., \& Shuder, J.~M., 1982, ApJS, 49, 149

\bibitem[]{} Pettini, M., Shapley, A.~E., Steidel, C.~C., Cuby, J.,
  Dickinson, M., Moorwood, A.~F.~M., Adelberger, K.~L., \& Giavalisco,
  M., 2001, ApJ, 554, 981

\bibitem[]{} Pope, A., et al.\, 2005, MNRAS, submitted.

\bibitem[]{} Ramsay Howat, S., Todd, S., Leggett, S., Davis,
  C,. Strachan, M., Borroeman, A., Ellis, M., Gostick, D., Kackley R.,
  Rippa, M., 2004, SPIE, in press.

\bibitem[]{} Scott, S.~E., Fox, M.~J., Dunlop, J.~S., Serjeant, S.;
  Peacock, J.~A., Ivison, R.~J., Oliver, S., Mann, R.~G., Lawrence, A.,
  Efstathiou, A., Rowan-Robinson, M., Hughes, D.~H., Archibald, E.~N.,
  Blain, A., Longair, M., 2002, MNRAS, 331, 817
  
\bibitem[]{} Shapley, A.~E., Steidel, C.~C., Pettini, M., Adelberger,
  K.~L., 2003, ApJ, 588, 65
  
\bibitem[]{} Shure, M., Toomey, D.~W., Rayner, J., Onaka, P., Denault,
  A., Stahlberger, W., Watanabe, D., Criez, K., Robertson, L., Cook, D.
  1993, AAS, 183, 11701

\bibitem[]{} Simpson, C., Rawlings, S., Lacy, M., 1999, MNRAS, 306, 828

\bibitem[]{} Smail, I., Ivison, R.~J., Blain, A.~W., 1997, ApJL,
  490, L5

\bibitem[]{} Smail, I., Ivison, R.~J., Blain, A.~W., Kneib, J.~P.,
  2002, MNRAS, 331, 495

\bibitem[]{} Smail, I., Chapman, S.~C., Ivison, R.~J., Blain, A.~W.,
  Takata, T., Heckman, T.~M., Dunlop, J.~S., Sekiguchi, K., 2003,
  MNRAS, 342, 1185
  
\bibitem[]{} Smail, I., Chapman, S.~C., Ivison, R.~J., Blain, A.~W.,
  2004, ApJ, 616, 71

\bibitem[]{} Steidel, C.~C., Adelberger, K.~L., Shapley, A.~E.,
  Pettini, M., Dickinson, M., Giavalisco, M., 2000, ApJ, 532, 170

\bibitem[]{} Storchi-Bergmann, T., Nemmen da Silva, R., Eracleous, M.,
  Halpern, J.~P., Wilson, A.~S., Filippenko, A.~V., Ruiz, M.~T., Smith,
  R.~C., Nagar, N.~M., 2003, ApJ, 598, 956

\bibitem[]{} Swinbank, A.~M., Smail, I., Chapman, S.~C., Ivison, R.~J.,
  \& Blain, A.~W., Keel, W., 2004, ApJ, 617, 67

\bibitem[]{} Teplitz, H.~I., McLean, I.~S., Becklin, E. E., Figer,
  D.~F., Gilbert, A.~M., Graham, J.~R., Larkin, J.~E., Levenson, N.~A.,
  Wilcox, M.~K., 2000, ApJ, 533, 65
  
\bibitem[]{} Wilman, R., Bower, R.~G., Morris, S.~L., Gerssen, J.,
  Bacon, R., Chapman, S., Davies, R.~L., de Zeeuw, P.~T., Emsellem, E.
  2005, in prep.

\bibitem[]{} Webb, T.~M.~A., Lilly, S.~J., Clements, D.~L., Eales, S.,
  Yun, M., Brodwin, M., Dunne, L., Gear, W.~K., 2003, ApJ, 597, 680

\bibitem[]{} White, S.~D.~M., Rees, M.~J., 1978, MNRAS, 183, 341
  
\bibitem[]{} Tecza, M., Baker, A.~J., Davies, R.~I., Genzel, R.,
  Lehnert, M.~D., Eisenhauer, F., Lutz, D., Nesvadba, N., Seitz, S.,
  Tacconi, L.~J., Thatte, N.~A., Abuter, R., Bender, R. 2004, ApJL,
  605, 109

\end{thebibliography}
\end{document}